\documentclass[fleqn,10pt]{article}
\usepackage{arxiv}
\usepackage{algorithm2e}
\usepackage{amssymb,amsthm,latexsym,amsmath,graphics,setspace, bm}
\usepackage{subfig}
\usepackage[superscript,biblabel]{cite}
\usepackage[thinlines]{easytable}
\usepackage{longtable}

\usepackage[utf8]{inputenc} 
\usepackage[T1]{fontenc}    
\usepackage{hyperref}       
\usepackage{url}            
\usepackage{booktabs}       
\usepackage{amsfonts}       
\usepackage{nicefrac}       
\usepackage{microtype}      
\usepackage{lipsum}
\usepackage{graphicx}
\usepackage{siunitx}
\usepackage{amsmath}
\usepackage{authblk}
\usepackage{accents}
\usepackage{mathtools}
\usepackage{authblk}

\newcommand{\bu}[1]{\mathbf{#1}}

\newcommand{\bb}[1]{\mathbb{#1}}

\title{Markov cohort state-transition model: A multinomial distribution representation}
\author[1,2,3]{Rowan Iskandar}
\author[1,2]{Cassandra Berns}

\affil[1]{Center of Excellence in Decision-Analytic Modeling and Health Economics Research,sitem-insel, Switzerland}
\affil[2]{Brown University, Providence, RI, USA}
\affil[3]{Institute of Social and Preventive Medicine, University of Bern, Switzerland}
\begin{document}
\maketitle
\begin{abstract}
    Markov cohort state-transition models have been the standard approach for simulating the prognosis of patients or, more generally, the life trajectories of individuals over a time period. 
    Current approaches for estimating the variance of a Markov model using a Monte Carlo sampling or a master equation representation are computationally expensive and analytically difficult to express and solve.
    We introduce an alternative representation of a Markov model in the form of a multinomial distribution.
    We derive this representation from principles and then verify its veracity in a simulation exercise.
    This representation provides an exact and fast approach to compute the variance and a way to estimate transition probabilities in a Bayesian setting.
\end{abstract}
\newpage
\section{Introduction}
Markov cohort state-transition models have been used in cost-effectiveness analyses and decision-analytic modeling studies to simulate the life trajectories of a patient or a group of patients following the receipt of a health intervention \cite{pauker1983}.  
Briefly, a Markov cohort state-transition model is a recursive matrix formula that calculates the \textit{average} number of individuals in each state using a transition probability matrix and a initial distribution of individuals across the states.  
This formulation introduced by Beck and Pauker \cite{pauker1983}, and its subsequent tutorial \cite{sonnenberg1993markov}, however, does not fully capture the \textit{stochastic} nature of the Markov model since it only represents the average behavior of the stochastic process \cite{iskandar2018theoretical}.
A more useful quantity is the probability distribution of all individuals at all times.
To obtain this quantity, Iskandar \cite{iskandar2018theoretical} explicated the stochastic process underlying a Markov model in the form of a well-known time-evolution equation of a probability function, i.e., a master equation, whose solution is the probability distribution of interest \cite{iskandar2018theoretical}.
Although the concept of a master equation and its solution provide a complete characterization of a Markov model, including higher-order moments, such a formulation requires some understanding of stochastic processes. 
This brief report introduces an alternative representation of a Markov model in the form of a multinomial distribution that is intuitive and hence more accessible to practitioners.
We start with a description of a Markov cohort state-transition model.
Then, we formally derive a multinomial distribution representation.
Lastly, we conduct a numerical exercise to verify its equivalence to a canonical approach for simulating cohort models. 
\section{Markov cohort state-transition model}
A cohort of $n_0$ individuals where each individual (indexed by $j=1,\ldots,n_0$) follows a Markov chain on a finite set of mutually exclusive and completely exhaustive $s$ health states: $\mathbb{S}=\{S_1,  S_2, \ldots, S_{s}\}$, is defined as a stochastic process, $\{\bu{N}(t)\}_{t\ge t_0}$, where $t$ and $t_0$ are time and the initial time, respectively.
$\bu{N}(t)$ represents a random vector of individuals across $s$ states (\textit{state-configuration}) :
\begin{align} \label{eq:trace}
    \mathbf{N}(t)=[N_1(t) \; N_2(t)  \; \ldots \; N_{s}(t)],
\end{align}
in which $\mathbf{N}(t)=\mathbf{n} \in \bb{N}^{s}$ ($\bb{N}$ is the set of non-negative integers).
The probability of observing a particular state-configuration $\mathbf{N}(t)=\mathbf{n} \in \bb{N}^{s}$ at time $t$ or $Pr[\bu{N}(t)=\bu{n}] \coloneqq G(\bu{n},t)$ (the notation $Pr[\cdot]$ denotes the probability of observing the random event $\{\cdot\}$ at time $t$) provides the complete information about the cohort model.
Our goal is to provide a multinomial distribution representation of this $Pr(\bu{N}(t)=\bu{n})$.
\section{Multinomial distribution representation}
To derive the multinomial representation, we start by deriving the stochastic process for each individual and aggregate the individual processes by appealing to the multinomial distribution.
\textbf{Individual process}. For each $j$-th individual, we define a $1 \times s$ random unit vector of health state occupancy, $\bu{Y}_j(t)$, where the value of the $i$-th element of $\bu{Y}_j(t)$, i.e., $Y_{j,i}(t)$, represents whether the individual is in state $S_i$: $1$ (occupied) or $0$ (not occupied), i.e., $\bu{Y}_j(t) \in \{0,1\}^s$ and $\sum\limits_{i=1}^{s}Y_{j,i}(t)=1$ (an individual must be in one and only one state at any time).
The dynamics of  $\bu{Y}_j(t)$ is governed by the allowed transitions at discrete time steps ($t_z$; $z=1,\ldots$) between all pairs of states in $\mathbb{S}$ and the intensities of these transitions.
We define an $s \times s$ stochastic matrix of transition probabilities governing the intensities of the transitions among states in $\mathbb{S}$, i.e., $\bu{P}_\tau(t_z)=\left[p_{kl}(t_z)\right]_{1\le k,l\le s}$, where $\tau$ denotes the time step (Markov cycle).
Each $p_{kl}(t)$ has the usual interpretation of the probability of an individual transitioning from states $S_k$ to $S_l$ in one time step $\tau$ (Markov cycle) at time $t_z$.
We are interested in deriving the probability that the individual $j$ occupies a particular state at any time $t_z$ given an initial condition. 
Such a probability is also a transition probability and can be derived by iteratively propagating the initial probability distribution of state occupancy (denoted by the $1 \times s$ vector $\bu{p}(t_z)$) at the initial time, i.e., $\bu{p}_j(t_0)$, forward in time using the transition probability matrix:
\begin{align}
    \bu{p}_j(t_z)=\bu{p}_j(t_0)\prod_{u=1}^{z-1}\bu{P}_{\tau}^j(t_u)
\end{align}
To derive the probability distribution of $\bu{Y}_j$, we note that there are $s$ possible realizations of $\bu{Y}_j$. 
The probability of a realization, i.e., the probability of an individual in some state $k$, is equal to $[\bu{p}_j(t_z)]_k$, where $[\cdot]_k$ denotes the $k$-th element of a vector (e.g., $[\bu{y}_{j}(t_Z)]_k=y_{j,k}(t_z)$).
We write probability distribution of $\bu{Y}_j(t_z)$ as follows:
\begin{align}\label{eq:pdf_multinomial}
    Pr\left(\bu{Y}_j(t_z)=\bu{y}_j(t_z)\right)=\prod\limits_{k=1}^{s}\left(\left[\bu{p}_j(t_z)\right]_{k}\right)^{[\bu{y}_{j}(t_Z)]_k}
\end{align}
Equation \ref{eq:pdf_multinomial} has the form of a multinomial distribution, where the number of individuals is equal to one ($n_0=1$).
Each term in the product corresponds to the probability of individual $j$ being in one of the $s$ states at time $t_z$.
\\ \\
The expected value of $\bu{Y}_j(t_z)$ is equal to $Pr[\bu{Y}_j(t_z)]$ since the expected value of each element of $\bu{Y}_j(t_z)$ is equal to the expected value of an indicator function of whether the $j$-th individual occupies the corresponding state, i.e., $E\left[Y_{j,k}(t_z)\right]=\left[\bu{p}_j(t_z)\right]_k=p_{j,k}(t_z)$.
The covariance matrix of  $\bu{Y}_j(t_z)$, denoted by $\Sigma_{\bu{Y_j}}(t_z)$, is calculated using the definition of a covariance matrix, $\left[\Sigma_{\bu{Y_j}}(t_z)\right]_{kl}=E[Y_{j,k}Y_{j,l}]-E[Y_{j,k}]E[Y_{j,l}]$, and is equal to:
\begin{align} \label{eq:individual_variance}
    \left[\Sigma_{\bu{Y_j}}(t_z)\right]_{kl} = \begin{cases}
     p_{j,k}(t_z)(1-p_{j,l}(t_z))  &\mbox{ if } k=l \\
   - p_{j,k}(t_z) p_{j,l}(t_z) &\mbox{ if } k\neq l
    \end{cases}
\end{align}
\textbf{From individuals to a cohort}. We relate $\bu{N}(t)$ (Equation \ref{eq:trace}) with $\bu{Y}_j(t)$ by using the following relationship:
\begin{equation} \label{eq:sum_relation}
    \bu{N}(t) = \sum\limits_{j=1}^{n_0} \bu{Y}_j(t)
\end{equation}
where we write a cohort as the sum of its individual members.
To derive the probability distribution of $\bu{N}(t)$, we need to associate each element in $\bu{N}(t)$ with the corresponding element in the sum.
Each element in the sum consists of a vector of $1$s and $0$s, i.e.,  $\bu{Y}_j(t)$.
The sum is constrained to be $n_0$ (a closed cohort).
The number of possible ways that $\bu{N}(t)$ can be realized, given that ordering of individuals within group does not matter, is equal to the number of possible realizations of the vector  $[\bu{Y}_1(t) \bu{Y}_2(t) \ldots \bu{Y}_{n_0}(t)]$.
Since all individuals are independent and follow the same transition rules and intensities ($\bu{P}_\tau^j(t)=\bu{P}_\tau(t)$, $p_{j,k}(t_z)=p_{k}(t_z)=$ for all $k$s, and $\bu{p}_j(t_0)=\bu{p}(t_0)$), each realization of $[\bu{Y}_1(t_z) \bu{Y}_2(t_z) \ldots \bu{Y}_{n_0}(t_z)]$ occurs with a probability of
\begin{align} \label{eq:prob_of_Ys}
 Pr([\bu{Y}_1(t_z) \bu{Y}_2(t_z) \ldots \bu{Y}_{n_0}(t_z)]) &=\prod\limits_{j=1}^{n_0}\prod\limits_{k=1}^{s}p_k(t_z)^{y_{j,k}(t)} \notag \\ 
  &= \prod\limits_{k=1}^{s}p_k(t_z)^{\sum\limits_{j=1}^{n_0}y_{j,k}(t)} \notag \\ 
    &= \prod\limits_{k=1}^{s}p_k(t_z)^{n_k(t)} 
\end{align}
\textbf{Multinomial distribution}. The number of ways that $\bu{N}(t)$ can be realized is identical to the problem of assigning $n_0$ of people into $s$ groups, which is given by the multinomial coefficient: $\binom{n_0}{n_1 n_2 \ldots n_s}$.
Putting all components together, the probability of $\bu{N}(t_z)$ at time $t_z$ is given by the following multinomial distribution:
\begin{align} \label{eq:multinomial}
    Pr(\bu{N}(t_z))=\binom{n_0}{n_1 n_2 \ldots n_s}\prod\limits_{k=1}^{s}p_k(t_z)^{n_k(t_z)} 
\end{align}
To derive the first two moments, we use the first two moments of the individual process ($\bu{Y}_j(t)$) and the relationship between each individual and a cohort of individuals (Equation \ref{eq:sum_relation}).
By linearity of expectation, the expected value of $\bu{N}(t_z)$ is equal to: \begin{align} \label{eq:multinomial_mean}
    E[\bu{N}(t_z)]=n_0 \bu{p}(t_z)
\end{align}
By independence of individuals, the variance of $\bu{N}(t_z)$ is the sum of $n_0$ identical variance of $\bu{Y_j}(t_z)$ and is given by:
\begin{align}
    \Sigma_{\bu{N}}(t_z)=n_0 \Sigma_{\bu{Y_j}}(t_z)
\end{align}
The covariance matrix of  $\bu{N}(t_z)$, denoted by $\Sigma_{\bu{N}}(t_z)$, is calculated using the definition of a covariance matrix, $\left[\Sigma_{\bu{N}}(t_z)\right]_{uv}=E[N_{u}N_{v}]-E[N_{u}]E[N_{u}]$, and is  equal to:
\begin{align} \label{eq:variance}
    \left[\Sigma_{\bu{N}}(t_z)\right]_{uv} = \begin{cases}
     n_0 p_u(t_z)(1-p_u(t_z))  &\mbox{ if } v=u \\
   -n_0 p_u(t_z) p_v(t_z) &\mbox{ if } v\neq u
    \end{cases}
\end{align}
\section{Numerical verification}
\begin{figure}
\centering
\includegraphics[angle=270,origin=c,width=\linewidth]{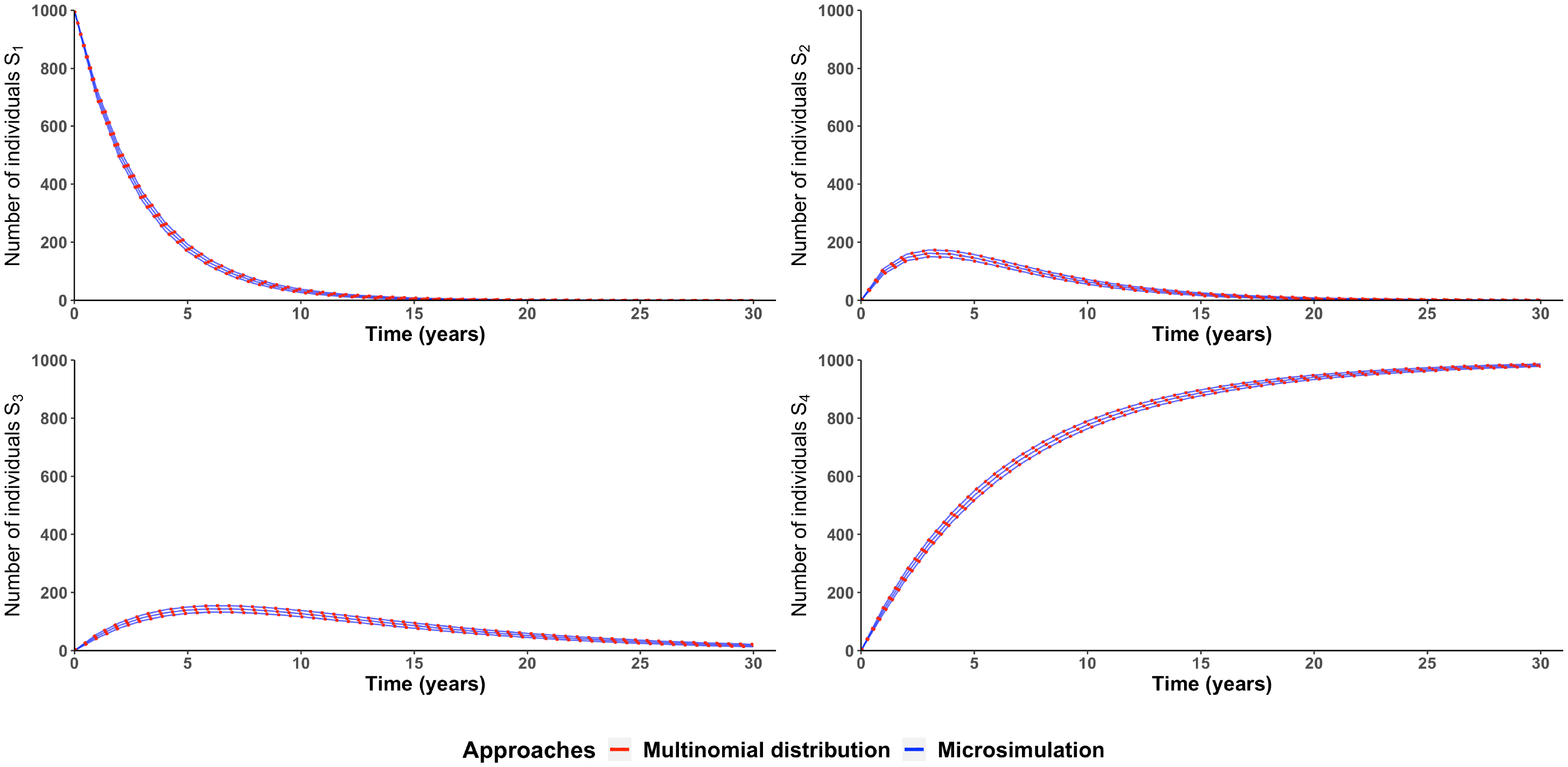}
\caption{The time trajectories of individual counts in each state using two approaches, i.e., microsimulation (blue solid lines) and multinomial distribution (red dotted lines). For each approach, there are three lines, i.e., mean, mean+standard deviation, mean-standard deviation. The lines are on the top of each other indicating no difference between the two approaches.}
\label{fig:figure XYZ}
\end{figure}

We conduct a simulation study to verify whether the multinomial distribution (Equation \ref{eq:multinomial}) represents the solution to a Markov cohort state-transition model.
We consider a $4$-state model with the allowed transitions and their probabilities as follows: $Pr[S_1 \rightarrow S_2]=0.1$,$Pr[S_1 \rightarrow S_3]=0.05$,$Pr[S_1 \rightarrow S_4]=0.14$, $Pr[S_2 \rightarrow S_3]=0.07$,$Pr[S_2 \rightarrow S_4]=0.17$, and $Pr[S_3 \rightarrow S_4]=0.11$. 
The simulation focuses on comparing the mean (Equation \ref{eq:multinomial_mean}) and variance (Equation \ref{eq:variance}) of the multinomial distribution with those of a microsimulation.
We simulate a cohort of 10000 individuals ($n_0=10000$) with $\tau=1$-year.
Each individual's life trajectory is a realization of a Markov chain based on the given transition probabilities.
We replicate the simulation $1000$ times and calculate the means and variances of the number of individuals across states at all times.
The results of the population trajectories across the four states are given in Figure \ref{fig:figure XYZ}.
We observe no difference between the empirical estimates of the mean and variance from the microsimulation and the multinomial distribution. 
The code for the numerical exercise is available under a GNU GPL license and can be found at https://github.com/rowaniskandar/CM\_multinomial.
\section{Concluding remarks}
This study explicates an alternative representation of a Markov cohort state-transition model in the form of a multinomial distribution.
We derive the equivalent representation by using elementary arguments.
First, the derivation starts with specifying a Markov state-transition model for simulating an individual following Beck and Pauker \cite{pauker1983}.
We then extend the model to a cohort of individuals by imposing independence among individuals and uncover the multinomial distribution.
We verify the first two moments of the derived multinomial distribution of the number of individuals across health states against using a microsimulation.
The formula for the first moment (Equation \ref{eq:multinomial_mean}) is indeed the formula for simulating Markov models as introduced by Beck and Pauker \cite{pauker1983}.
To estimate the variance, practitioners often use microsimulations, which may be computationally expensive since we need to replicate the cohort simulation a number of times in addition to the individual Monte Carlo runs within each simulated cohort.
In contrast, the multinomial representation provides a more direct and non-computationally demanding approach, particularly for estimating the variance. 
This approach relies only on the formula for the mean, which in turn depends only on the initial distribution and the time-dependent or time-invariant translation probability matrix.
In addition to the computational advantage, the multinomial representation provides a convenient way to conduct Bayesian inference on the transition probabilities \cite{walley1996inferences}.
In a Bayesian setting, the Markov model would be treated as the likelihood and naturally takes the form of a multinomial distribution using the result of this study.
We can then derive the posterior distribution of the transition probabilities by utilizing the conjugacy between the Dirichlet and multinomial distributions.
In sum, this study introduces another formulation of a well-established methodology and reinforces the utility of Markov models further.
\bibliography{CM_SDE_bib}
\bibliographystyle{unsrtnat}
\end{document}